# A PROTOTYPE COMPACT ACCELERATOR DRIVEN NEUTRON SOURCE FOR CANADA SUPPORTING MEDICAL AND SCIENTIFIC APPLICATIONS

**Dalini D. Maharaj,**[a,b] **Mina Abbaslou,**[b,c] **Sana Tabbassum,**[d] **Alexander Gottberg,**[b,c] **Marco Marchetto,**[b] **Zin Tun,**[e] **Linda H. Nie,**[d] **Oliver Kester,**[b,c] **Drew Marquardt,**[a] **and Robert Laxdal**[b,c]

[a] Department of Chemistry and Biochemistry, University of Windsor, 401 Sunset Avenue, Windsor, Ontario N9B 3P4, Canada
[b] TRIUMF, 4004 Wesbrook Mall, Vancouver, British Columbia V6T 2A3, Canada
[c] Department of Physics and Astronomy, University of Victoria, PO Box 1700 STN CSC, Victoria, British Columbia V8W 2Y2, Canada
[d] School of Health Sciences, Purdue University, 550 Stadium Mall Drive, West Lafayette, Indiana 47907, United States of America
[e] TVB Associates Inc., 691 Upper Wentworth Street, Hamilton, Ontario L9A 4V6, Canada

**ABSTRACT**

Canada recently lost its major supply of neutron beams with the closure of the National Research Universal reactor at Chalk River Laboratories in March 2018. This issue is further exacerbated by the closure of other reactors abroad, which also served as centers for neutron research. Consequently, there is a demand for new sources, both for Canada and internationally, as the global supply shrinks. Compact accelerator driven neutron sources provide an avenue to realize an intense source of pulsed neutron beams, with a capital cost significantly lower than spallation sources. In an effort to close the neutron gap in Canada, a Prototype Canadian compact accelerator driven neutron source (PC-CANS) is proposed for installation at the University of Windsor. The PC-CANS is envisaged to serve two neutron science instruments, a boron neutron capture therapy (BNCT) station and a beamline for fluorine-18 radioisotope production for positive emission tomography. To serve these diverse applications of neutron beams, a linear accelerator solution is selected to provide 10 MeV protons with a peak current of 10 mA within a 5% duty cycle. The accelerator is based on an RFQ and a DTL with a post-DTL pulsed kicker system to simultaneously deliver macro-pulses to each end-station. The neutron production targets for both neutron science and BNCT will be made of beryllium and engineered to handle the high beam power density. Conceptual studies of the accelerator and benchmarking studies of neutron production and moderation with FLUKA and MCNP, in support of the target-moderator-reflector design are reported.

**KEYWORDS**
Linear Accelerator, Neutron Source, Monte Carlo

# 1. INTRODUCTION

For over 75 years, neutron beams have been utilized as a novel probe of matter in a diversity of fields including condensed matter physics, biological sciences, geology, cultural heritage, and medicine [1]. Canada is currently facing a drought of neutron beams with the closure of the National Research Universal reactor at Chalk River Laboratories and curtailment of the scientific user program in March 2018 [2]. The global neutron community is facing a similar fate with the closure of reactors around the world [3]. While new generation spallation neutron sources like the Spallation Neutron Source (SNS) [4] and the China Spallation Neutron Source (CSNS) [5] provide pulsed, high-brilliance neutron beams, they are costly and will be unable to meet the growing demand for neutron beams.

In an effort to address this challenge, the global neutron community is responding by taking advantage of recent advances in accelerator technology to develop compact accelerator driven neutron sources (CANS). CANS generate neutrons via the bombardment of light metal targets, for example lithium (Li) or beryllium (Be), with pulsed proton beams with high peak-current and energies on the order of 2.5-10 MeV [6]. CANS can provide a local neutron source at a fraction of the cost of a spallation source (or reactor) given the 100-fold reduction in the required energy. The lower energy also reduces the required shielding and allows neutron instruments to be placed closer to the source, permitting an efficient transfer of the neutron phase space volume. The source can also be highly optimized to meet the end requirements of each instrument, resulting in the production of brilliant beams to rival that of small to medium sized reactor sources.

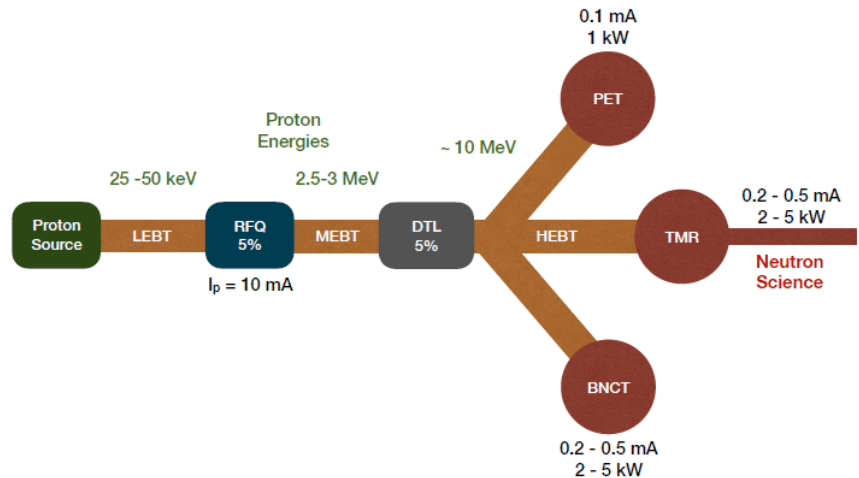

**Figure 1: The schematic of the PC-CANS is illustrated.**

The initiative to develop a prototype Canadian CANS (PC-CANS) at the University of Windsor seeks to close the neutron gap in Canada, demonstrate proof of concept and realize a medium brilliance neutron source. The PC-CANS will host three end-uses; neutron sciences, boron neutron capture therapy (BNCT) and fluorine-18 (F-18) medical isotope production for positive emission tomography (PET), as shown in Figure 1 [7]. The highest priority beamline for the PC-CANS is the neutron science beamline which will facilitate time-of-flight diffraction and small-angle neutron scattering (SANS) experiments. The neutron science instruments will be designed to yield a neutron brilliance a factor of five greater than the proposed NOVA ERA facility [8] with a goal to deliver neutron beams with brightness no less than a factor of five lower than a medium-brightness spallation source like the ISIS neutron facility in the United Kingdom [9]. In this paper, we provide an overview of the facility envisaged at the University of Windsor in Section 2, present the potential variants for the accelerator design in Section 3 and present important neutron yield and moderation benchmarking studies with FLUKA [10] and MCNP [11] in Section 4.

## 2. OVERVIEW OF THE PC-CANS

The motivation and parameter definition of the PC-CANS has been previously presented [7]. Our plan is to target a facility in the range ~C$10M with a performance that would produce five times the cold neutron flux as that achievable at the SANS instrument proposed in the NOVA ERA conceptual design report [8], while supplying R&D levels of neutrons for BNCT and competitive yields of F-18 for PET. NOVA ERA is designed to produce 10 MeV protons at a peak current of 1 mA at 4% duty factor, leading to an average power on target of 0.4 kW [8]. PC-CANS also adopts a proton energy of 10 MeV but the average proton current is increased to 200 µA/2 kW for the neutron science target. The technology of the target within the target-moderator-reflector (TMR) system is viewed as the limiting factor defining the final average beam power. We choose to slightly increase the duty factor to 5% for a peak proton current of 4 mA/40 kW with repetition rates near 100 Hz. For PET isotope production 10 MeV protons will produce competitive yields of F-18 with 50 to 100 µA on target. The BNCT target is specified to receive an average current of 200 µA with 5% duty factor at 10 MeV (2 kW) and is expected to provide an epithermal neutron yield of $1.0 \times 10^8$ n/s/cm$^2$. While this is ~10 times less than required to produce therapeutic epithermal beams for BNCT treatments, the BNCT station will provide an entry level facility for pre-clinical research that allows a development path for a future BNCT machine. In summary, the neutron science and BNCT stations are specified for average currents of 200 µA each while the F-18 station is specified for an average current of 100 µA.

The peak intensity from the linear accelerator (or, linac) is specified to allow achieving these beam intensities assuming a simultaneous time share between the end stations, so that a total peak intensity of 10 mA at 5% duty factor is foreseen. This corresponds to peak currents of 4 mA, 4 mA and 2 mA being delivered to each of the neutron science, BNCT and F-18 targets, respectively. Beyond this, to add further upgrade potential and engineering margin, the linac is being designed for twice this current or a maximum peak intensity of 20 mA, corresponding to a peak power of 200 kW. These values are summarized in Table I. As target technology evolves the full average proton intensity, 500 µA, could be sent to the neutron science target for an average yield of 12.5 times achievable at NOVA ERA. Finally, the linac design is compatible with a total average current of 1 mA such that, if the target engineering evolves, an average neutron yield of 25 times NOVA ERA source could be reached. Since the protons are pulsed at 5% duty factor, the peak neutron flux is a factor of 20 higher than from the same continuous-wave (CW) machine demonstrating the key advantage, for example, over a 1 mA CW cyclotron.

Table I: Summary of PC-CANS requirements compared to the NOVA ERA design.

| Station | Energy (MeV) | $I_{av}$ (µA) | DF (%) | $P_{av}$ (kW) | $I_{pk}$ (mA) | $P_{pk}$ (kW) |
|---|---|---|---|---|---|---|
| NOVA ERA | 10 | 40 | 4 | 0.4 | 1 | 10 |
| PC-CANS | | | | | | |
| Neutron | 10 | 200 | 5 | 2 | 4 | 40 |
| F-18 | 10 | 100 | 5 | 1 | 2 | 20 |
| BNCT | 10 | 200 | 5 | 2 | 4 | 40 |
| Target Totals | | 500 | 5 | 5 | 10 | 100 |
| Linac Design | 10 | 1000 | 5 | 10 | 20 | 200 |

## 3. ACCELERATOR DESIGN

A design requirement of the linac is the need to operate the PC-CANS in a non-laboratory setting. This favors robust technology and simplified tuning requirements. This requirement and the pulsed low duty cycle scheme for the PC-CANS favors a normal conducting linac. In general, normal conducting cavities are more cost favorable than superconducting technology for low energy, high beam current and low duty cycle applications, especially for a non-accelerator laboratory setting. The typical layout of a modern proton linac comprises an Electron Cyclotron Resonance Ion Source (ECRIS), a Low Energy Beam Transport section (LEBT), a Radio Frequency Quadrupole (RFQ), a Medium Energy Beam Transport section (MEBT), a Drift Tube Linac (DTL) and a high energy beam transport (HEBT). The main considerations in choosing the technology are capital and operating costs, technical risk, reliability and availability.

Recent years have seen a significant progress in the development of a new generation of optimized linac systems and structures supported by large scale projects like SNS [4], CSNS [5], Japan Proton Accelerator Research Complex (J-PARC) [14], European Spallation Source (ESS) [13] and CERN [14]. Based on these projects, the PC-CANS strawman design assumes a source energy of 45 kV (in the range of commercial sources), an RFQ output energy of 3 MeV and a DTL further accelerating the protons to 10 MeV with a common rf frequency of 352 MHz. Preliminary beam dynamics calculations have been done for the RFQ and DTL variants. The RFQ baseline vane parameters and beam simulations have been realized using PARMTEQ [15]. A two solenoid LEBT is used to match the 20 mA beam into the RFQ. The PARMTEQ output for the current RFQ design is shown in Figure 2. The simulated transmission is 97.7% for a beam intensity of 20 mA using 295 cells for an RFQ length of 3270 mm. Initial normalized beam emittances of 0.25 µm rms have been used with a resulting rms emittance growth of < 10% and longitudinal emittance of 0.137 MeV-deg (at 352 MHz).

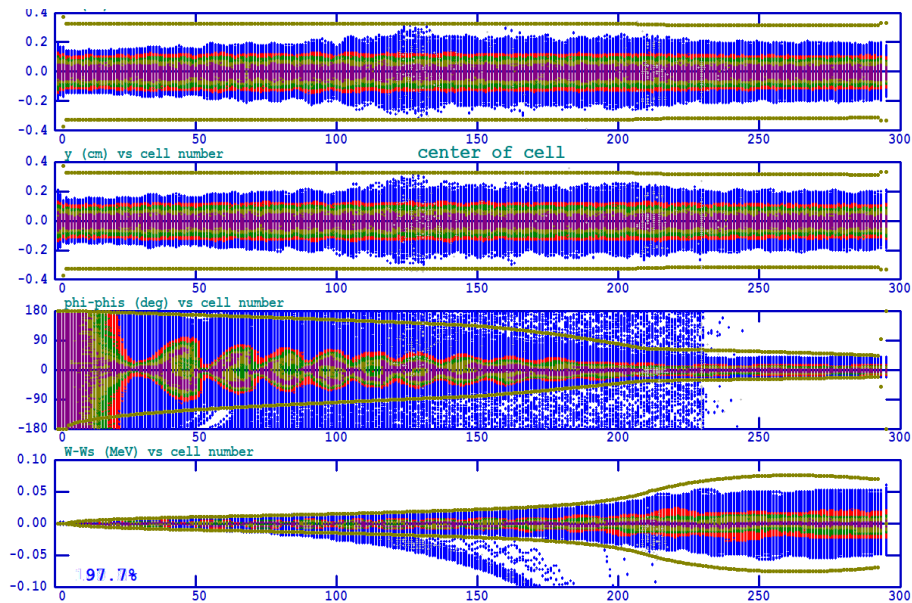

**Figure 2: PARMTEQ simulation of the RFQ shows the multiparticle coordinates as a function of cell number. The graphs from top to bottom give the x position (cm), y position (cm), phase (deg) and W-Ws (MeV).**

Two types of DTL have been studied: an Alvarez structure and a cross-bar H-mode (CH-DTL) structure. The Alvarez operates in TM010 mode with $2\pi$ phase shift between accelerating gaps. The drift tubes host either electro-magnetic quadrupole (EMQ) or permanent magnet quadrupoles (PMQ) to provide strong transverse focus during acceleration typically in a FODO lattice. The CH-DTL utilizes the TE210 mode that can be used to generate on axis accelerating fields through crossbar stems connected to drift tubes with $\pi$ phase shift between drift tubes. The H-mode structures are characterized by small drift tubes with no transverse focusing yielding a higher shunt impedance compared to the Alvarez, allowing a more aggressive accelerating gradient for the same peak rf power. In this energy and frequency range the expected shunt impedance for CH-structures and Alvarez structures are 80 MΩ/m and 45 MΩ/m respectively [16]. Conversely, the lack of transverse focussing in the CH-variant necessitates the use of shorter accelerating sections with quadrupole triplets or doublets in between the tanks or incorporated in the rf-tanks periodically to maintain transverse focussing. The Alvarez variant with built in optics has less 'knobs' and should be more straight-forward to tune while the CH variants allow some level of flexibility in the final energy and allows to apply the KONUS beam dynamics [17].

An Alvarez structure and three CH-DTL variants are being evaluated in terms of beam dynamics using multiple particle codes PARMILA [18] and LANA [19]. Trace-3D [20] is used to make a first estimate of matching parameters and number of cells. The variants are chosen to compare and contrast typical solutions that have been proposed for other applications. A summary of the four cases is presented in Figure 3 scaled for length where the PARMTEQ RFQ output beam parameters with 5xRMS emittance values and a beam intensity of 20 mA are used as the input in all cases. In most cases a four quadrupole MEBT of length 840 mm with a two gap 352 MHz rebuncher has been used to match from the RFQ to the DTL variants.

In the case of the Alvarez, an option with (Variant 1) and without MEBT (Variant 2) is considered since the strong transverse focussing in the drift tubes can be tuned to match the RFQ beam to the downstream FODO lattice. In the one variant shown in Figure 3, variable strength quadrupoles (EMQ) in the first four drift tubes are used to match the beam transversally with -30 degree synchronous phase adopted for longitudinal focussing. In the FODO section every second drift tube has a PMQ, with alternate drift tubes empty that could potentially be used for steering. The Alvarez tank consists of 26 drift tubes with 4 EMQs and 11 PMQs at ±64 T/m with an accelerating gradient of 3.2 MV/m. A CH variant (Variant 3) with KONUS (0 degree) beam dynamics consists of two tanks with 12 and 15 drift tubes in each tank and a quadrupole triplet of length 37 cm between tanks. At a gradient of 6 MV/m the estimated rf power consumption and length of Variant 3 is the same as the Alvarez Variant 2. Variant 4 uses traditional beam dynamics with a constant negative synchronous phase. The added transverse rf defocusing requires at least three tanks and two sets of triplets with 13, 10 and 14 cells respectively with an assumed gradient of 4.5 MV/m. The last CH variant (Variant 5) is composed of five short tanks with 5, 5, 6, 7 and 7 drift tubes in each tank with 24 cm long doublets providing focussing between tanks. Variant 5 would reduce the cost of an rf unit but would require more rf units and the shorter doublet focussing sections would give less variation in the phase spread of the longitudinal phase space.

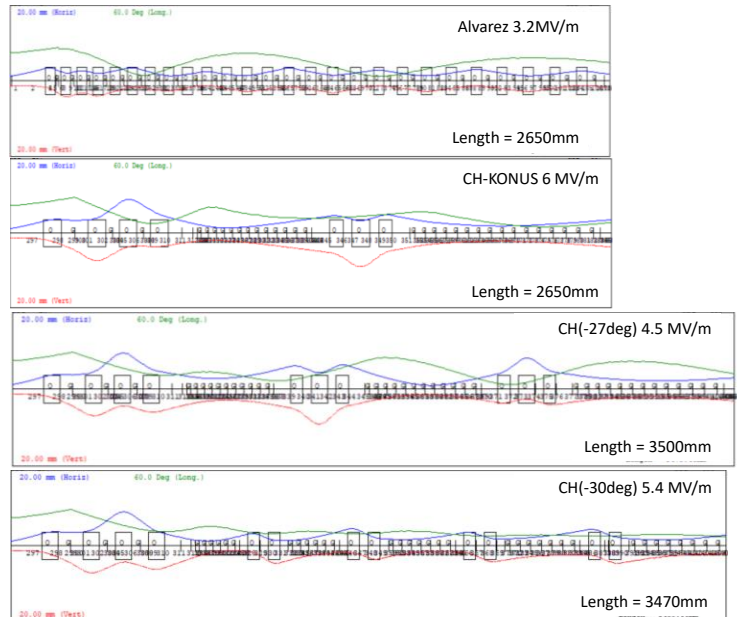

Figure 3: The x (blue), y (red), z (green) envelopes of the four DTL variants scaled for length. All simulations assume the output Twiss parameters from the PARMTEQ RFQ simulation with 5xRMS emittances.

Multi-particle simulations, sensitivity studies, rf simulations and cost comparisons are being done for all variants to facilitate a down select to the final proposal. Early studies have been done to benchmark LANA and PARMILA using the Alvarez no-MEBT variant as a test case. The output phase space from the two codes are presented in Figure 4, where the input particle distributions are derived from the PARMTEQ exit Twiss parameters assuming 5xRMS emittances. The results compare well.

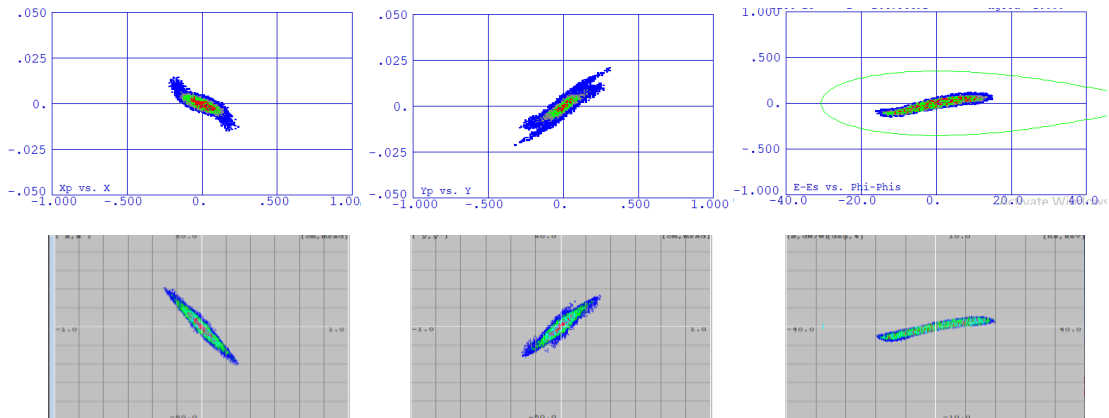

**Figure 4: Phase space plots (from left to right x:x´, y:y´, ΔE:φ) at the end of the Alvarez DTL ''no MEBT'' variant from PARMILA (top) and LANA (bottom) plotted on the same scale. The simulations each assume the 5RMS PARMTEQ emittance values at 20 mA beam intensity.**

A pulsed switchyard downstream of the linac in the HEBT can be used to direct three beams simultaneously into the three different target stations. An example of a particular solution for a pulsed switchyard design is shown in Figure 5. A beam with a pulse repetition rate of 200 Hz (5 ms period, 250 µs pulse length) and an average current of 500 µA is first deflected by a kicker at 50 Hz, deflecting every fourth macro-pulse toward the PET beamline with an average current of 125 µA. The rest of the beam is transported to a second kicker at 100 Hz and phased to deflect every second pulse of the initial beam into the neutron science beamline for an average current of 250 µA and a macro-pulse period of 10 ms suitable for time-of-flight studies. The undeflected pulsed beam with a current of 125 µA that remains is directed to the BNCT station.

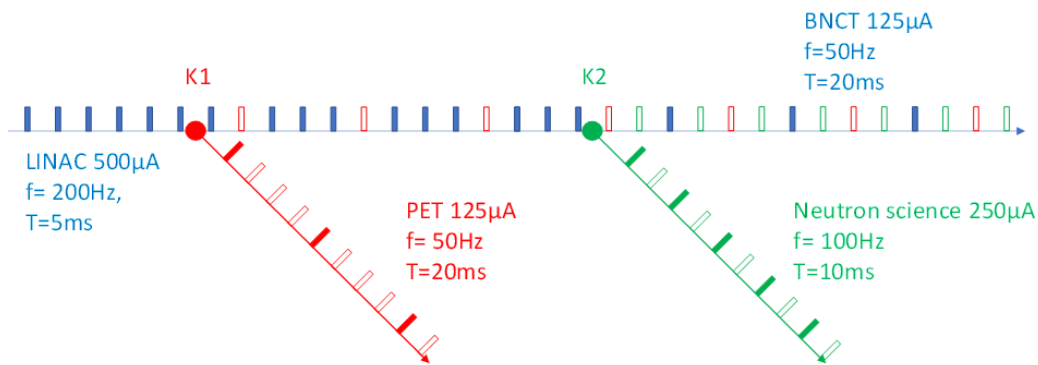

**Figure 5: A pulsed beam with the frequency of 200 Hz, pulse length of 250 µs, duty cycle of 5%, and current of 500 µA split into three target stations by a 50 Hz kicker (K1) and a 100 Hz kicker (K2); f is the macro-pulse frequency and T is the macro-pulse period.**

## 4. MONTE-CARLO SIMULATIONS

### 4.1. Proton-to-Neutron Conversion

In this section, we report preliminary work to benchmark the neutron yields from the $^9$Be(p,xn)$^9$B stripping reaction as calculated by FLUKA and MCNP in the proton energy range 3-10 MeV. Simulations of the total neutron yield (unmoderated) as a function of neutron energy from a thin beryllium target with proton energies of 3 MeV, 5 MeV, 7 MeV and 10 MeV are presented in Figure 6. In this comparison, the energy bins which are utilized to obtain the neutron spectrum in MCNP coincide with the 260 low-energy neutron transport groups which FLUKA uses to simulate the transport of neutrons with energies below 20 MeV. The range of 10 MeV protons in beryllium was determined to be 0.7 mm with the use of SRIM [21]. In these simulations, a spherical beryllium target of diameter, 0.7 mm, was employed to (1) maximize the yield from the proton interacting with the Be target and (2) limit the effect of neutron moderation by Be. Figure 6e.) reflects that agreement between FLUKA and MCNP for unmoderated neutron yields improves as the incident proton beam energy increases and at 10 MeV the difference between the total neutron yields is within 5%. Features of the spectra in Figure 6 a.) - d.) are reasonably matched with good agreement at 10 MeV. Table II compares the total neutron yields in MCNP and FLUKA. The uncertainties reflected in Table II are obtained from the statistics employed in the Monte Carlo simulations and not from uncertainties in the cross sections, hence, the true uncertainty is larger than reported in this table. These results reflect that MCNP and FLUKA can be reliably utilized to conduct side-by-side comparisons when simulating neutron production with 10 MeV protons on a Be target.

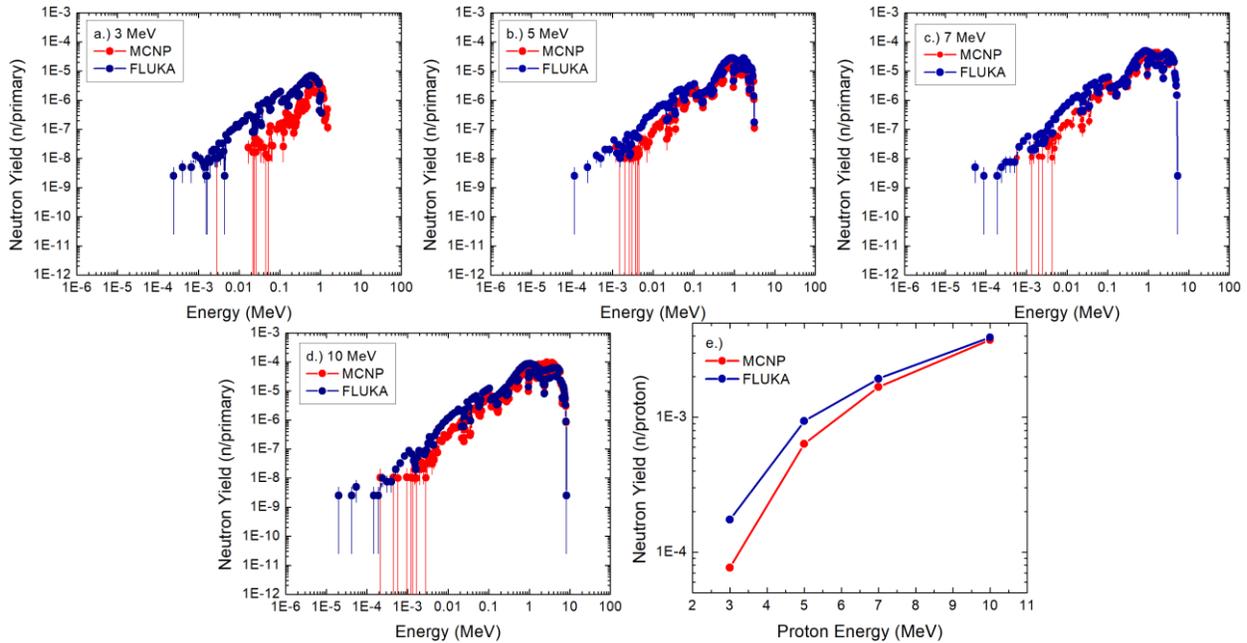

**Figure 6:** FLUKA and MCNP results showing the neutron spectrum for a) 3 MeV, b) 5 MeV, c) 7 MeV and d) 10 MeV protons impinging on a thin beryllium target of thickness, 0.7mm. The total neutron yield per proton as a function of proton energy is shown in panel e).

**Table II: The total neutron yields obtained from MCNP and FLUKA are presented.**

| Proton Energy (MeV) | MCNP | | FLUKA | |
|---|---|---|---|---|
| | Neutron Yield (n/primary) | Absolute Error (n/primary) | Neutron Yield (n/primary) | Absolute Error (n/primary) |
| 3 | 7.67E-5 | 1E-6 | 1.74E-4 | 7E-7 |
| 5 | 6.34E-4 | 2E-6 | 9.37E-4 | 2E-6 |
| 7 | 1.68E-3 | 5E-6 | 1. 93E-3 | 2E-6 |
| 10 | 3.74E-3 | 6E-6 | 3. 92E-3 | 3E-6 |

### 4.2. Neutron Moderation

Motivated by our previous work [7], we seek to benchmark the performance of MCNP and FLUKA for simulating neutron moderation using common moderator materials including, water, methane, polyethylene and parahydrogen. The geometry utilized for the simulations is shown in Figure 7. The simulations assume 10 MeV protons, propagating in the $+z$ direction and striking a spherical beryllium target of diameter 2 mm, which is centered at $(x,y,z) = (0,0,0)$ within a spherical moderator of 50 cm radius. The detectors are arranged in three directions along the $\pm z$ and $+x$ axes and are centered at the positions $z = \pm 5, \pm 10, \pm 15, \pm 20, \pm 25$, and $\pm 30$ cm and $x = +5, +10, +15, +20, +25$, and $+30$ cm in order to investigate the azimuthal dependence. In FLUKA, the densities of water, parahydrogen, methane and polyethylene are 1.0, 0.0708, 0.42 and 0.94 g-cm$^{-3}$ respectively. For parahydrogen and methane, the cross sections of hydrogen and carbon at 86 K are employed as lower temperature cross sections are unavailable. In the MCNP simulations, the cross sections at 20 K are utilized for parahydrogen and methane. In the FLUKA simulations, the differential neutron flux is calculated using the USRBDX card by selecting the scoring routine, $\Phi 1,\log(E),\lin(\Omega)$, in order to score the neutron fluence and one solid-angle bin is selected. The average surface flux tally (F2), which also calculates the neutron fluence is used in the MCNP simulations. The neutron spectra obtained from FLUKA are automatically assigned low-energy neutron groups as FLUKA performs low-energy neutron transport with groupwise cross sections while MCNP uses pointwise cross sections.

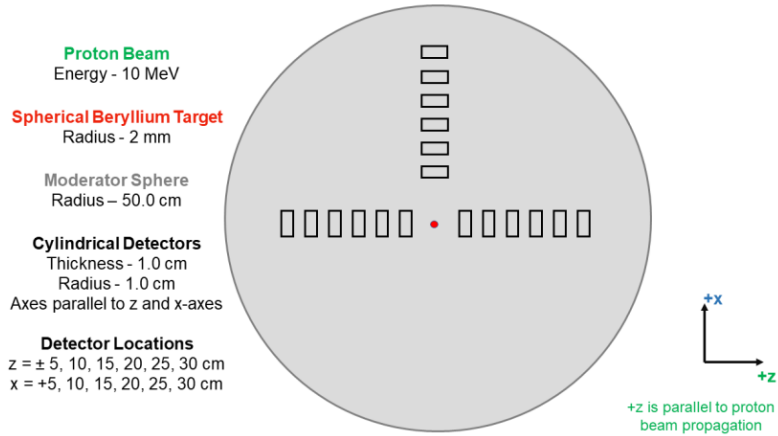

**Figure 7: The geometry utilized for the moderation study is illustrated.**

The neutron spectra obtained from the detectors positioned 5 cm away from the target and in the $+z$, $-z$ and $+x$ directions for the water moderator are shown in Figures 8 a.), b.) and c.), respectively. In the inset of each graph, the total neutron flux, integrated over the entire range of the spectrum is presented as a function of distance from the target center. In general, features of the neutron spectra as simulated in MCNP and

FLUKA are in excellent agreement. The total neutron flux from MCNP is systematically higher across the energy band regardless of the distance from the target by factors of 1.2-2.4 and the difference shows little azimuthal dependence. Figure 8 d.) presents the result for the detector at $+z = 5$ cm in the polyethylene case, where the total neutron flux in MCNP exceeds that of FLUKA by a factor 1.0-1.8. Figure 8 e.) and f.) show the spectra obtained when methane and parahydrogen are employed as moderators. The spectra details in the cold/thermal regime are in slight variance compared to the non-cryogenic cases but can be explained by the fact that FLUKA utilizes cross sections for free-hydrogen and free-carbon at 87K while MCNP uses the cross sections at 20 K. In addition, MCNP includes the bonding character of the moderators in the cross section. Outside of the cold/thermal neutron energy range, the spectra match well. Our results highlight that when neutron moderation simulations are performed with water and polyethylene, FLUKA and MCNP can both be used reliably, and in parallel to perform TMR optimization studies. However, when TMR optimization studies are conducted with cryogenic materials MCNP may provide more accurate results when compared with FLUKA.

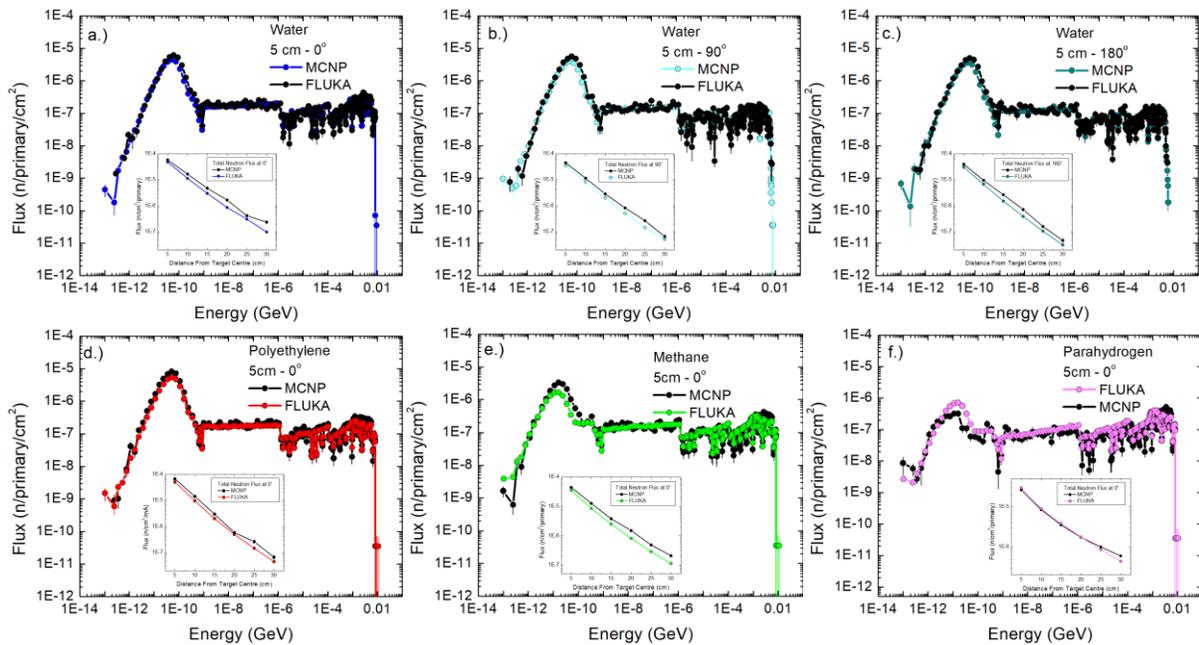

**Figure 8: The azimuthal dependence of the neutron spectrum with a water moderator is shown in panels a.) forward (0°), b.) perpendicular (90°) and c.) backward (180°). Neutron spectra obtained from the detector at z = 5cm are shown in d.) polyethylene, e.) methane and f.) parahydrogen. The inset plots illustrate the total neutron flux as a function of distance.**

## 5. CONCLUSIONS

PC-CANS represents a significant opportunity for the Canadian neutron user community while delivering PET isotopes and R&D opportunities for BNCT studies. A proposal is being prepared for a Canadian funding opportunity in early 2022. Presently various linac variants are being modeled and costed in preparation for the proposal. TMR designs have been aided by studies that benchmark FLUKA and MCNP to give confidence in the neutron yields as a function of various TMR geometries.